\documentclass[10pt,preprint]{revtex4}
\usepackage{graphicx}
\usepackage{amsmath}
\usepackage{amssymb}

\begin{document}

\title{Five-dimensional Monopole Equation 
 with Hedge-Hog Ansatz\\ and Abel's Differential Equation}
\author{Hironobu Kihara}
\affiliation{Korea Institute for Advanced Study\\
207-43 Cheongnyangni 2-dong, Dongdaemun-gu, Seoul 130-722, Republic of Korea}
\date{February 22, 2008}
\preprint{OCU-PHYS-291}
\preprint{KIAS-P08018}

\begin{abstract}
We review the generalized monopole in the five-dimensional Euclidean space. 
A numerical solution with the Hedge-Hog ansatz is studied. 
The Bogomol'nyi equation becomes a second order autonomous non-linear differential equation. 
The equation can be translated into the Abel's differential equation of the second kind and is an algebraic differential equation. 
\end{abstract}

\maketitle

\section{Tchrakian monopole in five dimensional space}
The Dirac's magnetic monopole has charmed many people for sixty years \cite{Dirac:1948um}.  
There are a lot of papers on the magnetic monopole though nobody find it up to now. 
The Dirac's monopole is described as a configuration of gauge potential which is singular at a point. 
Because of the singularity, its mass or energy cannot be determined. 
't Hooft and Polyakov realized the monopole as a classical solution of a non-Abelian gauge theory with adjoint scalar fields \cite{'t Hooft:1974qc,Polyakov:1974ek}. Their solution has finite energy. 

Yang considered the generalization of the Dirac monopole in five-dimensional space \cite{Yang:1977qv} and 
Tchrakian \cite{Tchrakian:1978sf} found various generalizations of the 't Hooft-Polyakov monopole and the Belavin-Polyakov-Schwartz-Tyupkin instanton \cite{Belavin:1975fg}. 

In this article, we focus on the five-dimensional monopole in one of the Tchrakian's models.  
The monopole is a classical configuration of the $SO(5)$ five-dimensional gauge theory with scalar fields which form a vector representation of $SO(5)$ with quartic terms with respect to the field strength. 
We need quartic terms to make the Bogomol'nyi completion \cite{Bogomolny:1975de}. 
In order to stabilize the asymptotic behavior of scalar fields, we only consider models which have the Higgs potential. 
We just consider equations after taking the Prasad-Sommerfield limit \cite{Prasad:1975kr}. 
We studied a Hedge-Hog solution of this model numerically \cite{Kihara:2004yz}. 
The Bogomol'nyi equations of this model are non-linear ordinary equations. 
These equations are reduced to one second-order equation. 
The equation is autonomous and we obtain a first-order equation. 
This equation is the Abel's ordinary differential equation of the second kind \cite{abel:1829,Murphy:1960}.

Let us consider the $SO(5)$ gauge theory with fundamental matter in five-dimensional Euclidean space, whose action or pseudo energy is
\begin{align*}
E &= \int  {\rm Tr} \left[  \frac{1}{8} F^{\wedge 2} \wedge * (F^{\wedge 2})  +\frac{1}{8}  D_A \phi \wedge *(D_A \phi) +\lambda  V(\phi)d^5x \right]~~.
\end{align*}  
Here $V(\phi)$ is the Higgs potential of scalar field and we take the Prasad-Sommerfield limit $\lambda \rightarrow 0$. The symbol ``$*$" represents the Hodge dual operator on the five-dimensional space with respect to the Euclidean metric. In this notation, we represent all multiplets in terms of the Clifford algebra which associates with the five-dimensional Euclidean metric. 
The Clifford algebra is generated by the Dirac's gamma matrices $\gamma_a, (a=1,2,3,4,5)$. They satisfy the anti-commutation relation: $\{ \gamma_a , \gamma_b \} = 2 \delta_{ab}$. 
These matrices are constructed by tensor products of the Pauli matrices $\sigma_a$ whose 
 multiplications is summarized in one equation: $\sigma_i \sigma_j = \delta_{ij} {\bf 1} + i \epsilon_{ijk} \sigma_k$ ($i,j,k=1,2,3$). 
Matrices $\gamma_1 = \sigma_1 \otimes 1, \gamma_2 = \sigma_2 \otimes 1, \gamma_3 = \sigma_3 \otimes \sigma_1, \gamma_4 = \sigma_3 \otimes \sigma_2, \gamma_5 = -\sigma_3 \otimes \sigma_3$  satisfy the anti-commutation relation and an additional relation $\gamma_1 \gamma_2 \gamma_3 \gamma_4 \gamma_5 =1$. Each $\gamma_a$ is a Hermitian matrix.
The generators of $SO(5)$ are represented as commutators of $\gamma_a$, $\gamma_{ab} = (1/2)[  \gamma_a , \gamma_b ] $. Let $k$ be an integer and ${\mathfrak S}_k$ be a symmetric group which consists of permutations of $k$ characters. Let us define the anti-symmetric product of gamma matrices: $\gamma_{m(1) \cdots m(k)} := (1/k!)\sum_{\tau \in {\mathfrak S}_k} \gamma_{m(\tau(1))} \cdots \gamma_{m(\tau(k))}$. 
We represent the gauge potential as $A= (1/2) A_m^{ab} \gamma_{ab} dx^m$ and the scalar field is also represented as $4\times 4$ matrices $\phi=\phi^a \gamma_a$. 
The field strength 2-form is given by $F=dA + g A^2$ where $g$ is the gauge coupling constant. 
A covariant derivative 1-form of $\phi$ is given by $D_A \phi = d \phi + g [A,\phi]$. 
The $N$-th power of a differential form $\omega$ is represented as $
\omega^{\wedge N}:= {\omega \wedge \omega \wedge \cdots \wedge \omega}$.
Let us use the notation $dx^{m_1m_2 \cdots m_n}=dx^{m_1} \wedge dx^{m_2} \wedge \cdots \wedge dx^{m_n}$.   All two elements in $\{ dx^{mn}\}$ commute with each other: $dx^{mn} \wedge dx^{pq} = dx^{pq} \wedge  dx^{mn}$. In order to explain the square of the field strength in terms of its  components, we show the anti-commutation relation of $\gamma_{ab}$: $\{ \gamma_{ab} , \gamma_{cd} \} = 2\gamma_{abcd} + 2(\delta_{bc}\delta_{ad} - \delta_{ac}\delta_{bd} )$. The relation $\gamma_1 \gamma_2 \gamma_3 \gamma_4 \gamma_5 = 1$ implies that $\gamma_{abcd} = \epsilon_{abcde} \gamma_e$, where $\epsilon_{abcde}$ is the five-dimensional Levi-Civita anti-symmetric tensor such that $\epsilon_{12345}=1$. 
The Kronecker delta on the rank-two anti-symmetric tensor is $\delta^{[ab]}_{cd}=(1/2)(\delta_{ac}\delta_{bd} - \delta_{ad}\delta_{bc})$. This tensor satisfies the relation $\delta^{[ab]}_{cd}T^{cd} = T^{ab}$ for all anti-symmetric tensor $T^{cd}$. 
Let us rewrite the pseudo-energy in terms of components. 
The square of the field strength $F =(1/4) F_{mn}^{ab} \gamma_{ab} dx^{mn}$ is 
$F \wedge F 
= (1/16) F_{mn}^{ab} F_{pq}^{cd} \left( \epsilon_{abcde}\gamma_e - 2 \delta^{[ab]}_{cd}    \right) dx^{mnpq}$. The components should be anti-symmetrized with respect to the spatial indices because the differential form $dx^{mnpq}$ is anti-symmetric. We define two four-rank anti-symmetric tensors $T_{mnpq}^e = (1/2\cdot 4!) \epsilon_{abcde} \left( F_{mn}^{ab} F_{pq}^{cd}+ F_{mp}^{ab} F_{qn}^{cd}+ F_{mq}^{ab} F_{np}^{cd} \right)$ and 
$S_{mnpq} = (1/ 4!)  \left( F_{mn}^{ab} F_{pq}^{ab}+ F_{mp}^{ab} F_{qn}^{ab}+ F_{mq}^{ab} F_{np}^{ab} \right)$. With these tensors, $F^{\wedge 2} = \left( T_{mnpq}^e \gamma_e - S_{mnpq} \right)dx^{mnpq}$. 
The gauge part of the density of the pseudo-energy is bilinear with respect to these tensors $T_{mnpq}^e, S_{mnpq}$:  $(1/4) {\rm Tr} F^{\wedge 2} \wedge * (F^{\wedge 2}) = \left( T_{mnpq}^e T_{mnpq}^e + S_{mnpq} S_{mnpq} \right) dv$ where $ dv$ is the volume form $dx^{12345}$.
The covariant derivative of the scalar field is $D_A\phi=(D_A\phi)_m^a \gamma_a dx^m$ and the pseudo-energy is 
\begin{align*}
E &= \int dv 
\left[ 
\frac{1}{2 \cdot 4!} \left( T_{mnpq}^e T_{mnpq}^e + S_{mnpq} S_{mnpq} \right)
+ \frac{1}{2} (D_A \phi)_m^a  (D_A \phi)_m^a  + \lambda V(\phi)
  \right]~~.
\end{align*}
The Bogomol'nyi completion from two terms $T^2$ and $D\phi^2$ yields a topological term. This  pseudo-energy is bound by the topological quantity $E \geq \int_{S^4} {\rm Tr}  \phi F \wedge F =Q$. 
Here $S^4$ is the sphere of radius $R$ which is large enough. We expect that the configuration localizes around a point and we take a limit $R \rightarrow + \infty$. 
For the field configuration which satisfy the Bogomol'nyi equation $\displaystyle F \wedge F = \pm * D_A \phi$, the pseudo-energy of this configuration attain the minimal quantity $Q$ after taking the Prasad-Sommerfield limit $\lambda \rightarrow 0$. 
Let $r^2=x^ax^a$ be the radius  and  $e=x^a \gamma_a/r$ be a unit vector. The solution of the Bogomol'nyi equation was studied with a hedgehog ansatz, $\phi = H_0 U(r) e$ and $A = (1-K(r))ede/2g$, numerically.
Here $H_0$ is the absolute value of vacuum configuration of the scalar field $\phi$. Let $\hat{x}_m = x_m/r$ be a unit vector of the radial direction. 
The differential form $ede$ is written in terms of these components; 
$ede = C_{mn}  \gamma_{mn}$ where $C_{mn} =(1/2) (\hat{x}_m d \hat{x}_n - \hat{x}_n d \hat{x}_m)$.  
 The expected boundary conditions are $U(0)=0,K(0)=1,U(\infty)=\pm 1, K(\infty)=0$. The covariant derivative $D_A\phi$ and the field strength are written in terms of $K$ and $U$:
 $D_A \phi = H_0 \left( U' e dr + KU de  \right)$, $F = (1-K^2)de \wedge de/4g - K' edr\wedge de/ 2g$. 
Similarly $F^{\wedge 2}$ is rewritten as $F \wedge F = \left({(1-K^2)}/{4g}  \right)^2 de^{\wedge 4} - ({(1-K^2)K'}/{4g^2}) edr \wedge de^{\wedge 3}$.
The Hodge dual of these forms can be read from 
$*\left( de^{\wedge 4} \right) = 4! edr/r^4~, * \left( edr \wedge de^{\wedge 3}   \right)= 3!de/r^2$.
The Bogomol'nyi equations reduce to a system of ordinary differential equations;
$
({4!}/{r^4}) \left( {(1-K^2)}/{4g}  \right)^2 = H_0 U'~, - ({3!}/{r^2}) 
{(1-K^2)K'}/{4g^2}=H_0KU$. 
We scale the radial coordinate $r$ by a factor $\displaystyle a = \left( {2g^2 H_0}/{3} \right)^{1/3}~, \tau = ar~$. 
\begin{align}
\frac{dU}{d\tau} &= \frac{1}{\tau^4} \left( 1-K^2 \right)^2~,& - \frac{1}{\tau^2} (1-K^2)\frac{dK}{d\tau} &= KU~.
\label{eqn:diff1}
\end{align}
Let us study the behavior of functions $K,U$ around boundaries. 
Let us consider the boundary $\tau=0$. 
Suppose that $K(\tau) = 1 + V(\tau)$. Then $V(\tau)$ and $U(\tau)$ are small enough around $\tau \sim 0$. We drop terms $V^3, VU$ and so on. 
From the equation (\ref{eqn:diff1}), We obtain $(d U/d\tau) = 4 V^2/\tau^4$, $(d V^2 /d \tau)/\tau^2 = U $. It implies that $U \sim \tau$ and $V \sim \tau^2$. 
Let us consider $\tau= \infty$. 
The differential $dU/d \tau$ should be positive and the initial value $U(0)$ is zero. It means that the preferable boundary condition is $U(\infty)=1$. Let us put $U=1 + W$. $W' \sim 1/\tau^4$, $- (1/\tau^2) K' \sim K$. Their solutions are $W \sim 1/\tau^3$ and $K \sim \exp ( - \tau^3/3)$. 
This system can be reduced to the first order ordinary equation. 
Let us change variables $s=\ln\tau$ $(- \infty < s < + \infty)$, $x(s) =K^2$ and let us eliminate the function $U$. 
Boundary conditions become $x( - \infty ) =1, x( + \infty) = 0$, $\dot{x}( - \infty ) = 0 , \dot{x}( + \infty ) = 0$, where we denote the derivative, $dx/ds$, ``$\dot{x}$". 
After that we obtain an autonomous equation; 
\begin{align}
\frac{d^2x}{ds^2} - \frac{1}{x(1-x)}\left( \frac{dx}{ds} \right)^2 -3 \frac{dx}{ds} + 2x(1-x)&=0~.
\label{eqn:2ndord}
\end{align}
This equation does not include the variable $s$ explicitly, ({\sl i.e.} it is autonomous) and we can reduce this equation to the first order differential equation of $y(x)=({dx}/{ds})/{x(1-x)}$. The boundary conditions  are 
$(x,y) = (1,-2)$ at $\tau \sim - \infty$ and $(x,y) = (0, \infty)$ at $\tau \sim + \infty$. 
\begin{figure}[tbh]
\begin{center}
\begin{minipage}{70mm}
         \includegraphics[width=60mm]{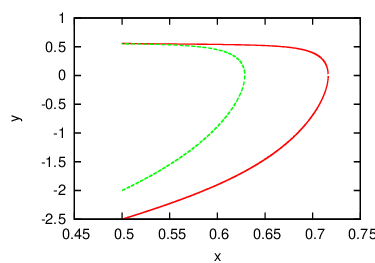}
      \caption{Movable branch point.}
      \label{fig:mov}
\end{minipage}~~
\begin{minipage}{70mm}
      \includegraphics[width=60mm]{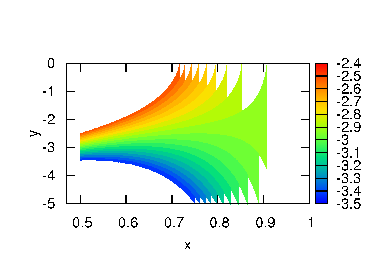}
      \caption{Flows near $(1,-2)$ in the $xy$-plane.}
      \label{fig:plusinf} 
\end{minipage}\\
\begin{minipage}{70mm}
         \includegraphics[width=60mm]{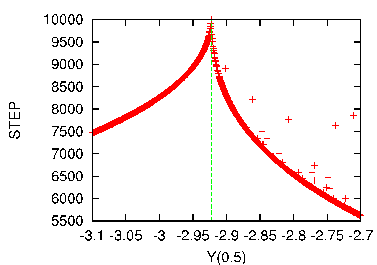}
      \caption{Numbers of steps in the Runge-Kutta method with thresholds (const.$<y<0$).}
      \label{fig:iv}
\end{minipage}~~
\begin{minipage}{70mm}
         \includegraphics[width=60mm]{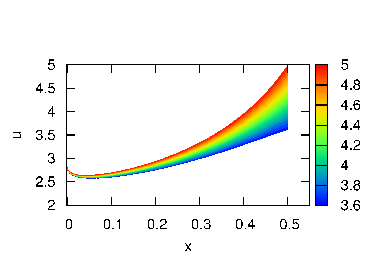}
      \caption{Flows near $(0,3)$ in the $xu$-plane.}
      \label{fig:minusinf}
\end{minipage}
    \end{center}
\end{figure}
\begin{align}
x(1-x)y \frac{dy}{dx} &= 2x y^2+3y-2~.
\label{eqn:abel1}
\end{align}
This equation (\ref{eqn:abel1}) is an Abel differential equation of the second kind \cite{abel:1829,Murphy:1960} and is an algebraic differential equation.  
This equation does not survive from the Painlev\'e's theorem and the equation might have a movable branch point (Fig.\ref{fig:mov}). 

Here we use the fourth order Runge-Kutta method and start from $x=1/2$. The initial value is  $y(1/2)$ in this case.  
The boundary $(1,-2)$ exists in finite range, while the boundary $(0,\infty)$ does not. 
Let us show flows around the boundary $(1,-2)$ (Fig.\ref{fig:plusinf}). 
Fig.\ref{fig:iv} shows that the initial value which gives a flow converging to the point $(1,-2)$ is $-2.9222$.   
In order to clarify the behavior around $(0,\infty)$ let us consider a transformation. 
Let us rewrite $y(x)$ in terms of $K,U,\tau$. 
From (\ref{eqn:diff1}), we obtain 
\begin{align*}
y(x) &= \frac{\tau dx/d\tau}{x(1-x)} = - \frac{2 \tau^3 U}{(1-K^2)^2}~.
\end{align*}
This implies that $y(x)$ behave like $\ln x$ around the boundary $\tau \sim + \infty$ because $K(\tau) \sim \exp(-\tau^3/3)$ and $U(\tau) \sim 1$ ($1<< \tau$). Therefore 
let us put $y(x)= (\ln x) u(x)$. 
\begin{align}
x(1-x)(\ln x) u \frac{du}{dx} &= \left\{ 2x(\ln x) - (1-x)  \right\} u^2 + {3u} - \frac{2}{\ln x} ~.
\label{eqn:ln}
\end{align}
Suppose that in the limit $x \rightarrow 0$, $u(x)$ and $(du/dx)(x)$ have finite limits;
$-{u^2} + {3u}  \rightarrow 0~,(\tau \rightarrow + \infty)~, u(+ \infty)= 0 ,3$.
In fact, Fig. \ref{fig:minusinf} shows that $u(+\infty) = 3$. 
This evaluation showed us the existence of the solution. 

Let us rewrite the differential equation (\ref{eqn:abel1}) in terms of the
 homogeneous coordinate $[X:Y:Z:W]$; $x=X/W,~y=Y/W,~(dy/dx)=Z/W$. Then they satisfy an algebraic relation $F(X,Y,Z,W) := X(W-X)YZ - 2XY^2W - 3 Y W^3 +2 W^4 =0$. This relation defines a projective variety in ${\mathbb C}{\bf P}^3$ and the degree is $4$. 
The variety is singular and it is not a K3 surface. 
We wonder if there exist another solutions which relate to K3 surfaces.

{\bf Acknowledgement}
I thank Yutaka Hosotani, Hiroshi Itoyama and Muneto Nitta for their teaching,  Tigran Tchrakian for his kind replies to my basic question, Yukinori Yasui, Ki-Myeong Lee, Piljin Yi for their advices, Kazuhito Fujiwara, Reiji Yoshioka, Kazunobu Maruyoshi, 
 Hyun Seok Yang, Kazuo Hosomichi, Qing-Guo Huang, Kyung-Kiu Kim, Sung-Jay Lee, and Eoin O Colgain for useful discussions. 

This work is partially supported by the 21 COE program ``Constitution of wide-angle mathematical basis focused on knots" of Osaka City University, Advanced Mathematical Institute. 

This work is supported by Korea Institute for Advanced Study founded by the Ministry of Science and Technology of Korea.




\begin{thebibliography}{99}

\bibitem{Dirac:1948um}
  P.~A.~M.~Dirac,
  Phys.\ Rev.\  {\bf 74} (1948) 817.

\bibitem{Polyakov:1974ek}
  A.~M.~Polyakov,
  JETP Lett.\  {\bf 20} (1974) 194
  [Pisma Zh.\ Eksp.\ Teor.\ Fiz.\  {\bf 20} (1974) 430].



\bibitem{'t Hooft:1974qc}
  G.~'t Hooft,
  Nucl.\ Phys.\  B {\bf 79}, 276 (1974).

\bibitem{Yang:1977qv}
  C.~N.~Yang,
  J.\ Math.\ Phys.\  {\bf 19}, 320 (1978).
\bibitem{Tchrakian:1978sf}
  D.~H.~Tchrakian,
  J.\ Math.\ Phys.\  {\bf 21}, 166 (1980).
  
\bibitem{Belavin:1975fg}
  A.~A.~Belavin, A.~M.~Polyakov, A.~S.~Schwartz and Yu.~S.~Tyupkin,
  Phys.\ Lett.\  B {\bf 59}, 85 (1975).





\bibitem{Bogomolny:1975de}
  E.~B.~Bogomol'nyi,
  Sov.\ J.\ Nucl.\ Phys.\  {\bf 24}, 449 (1976)
  [Yad.\ Fiz.\  {\bf 24}, 861 (1976)].

\bibitem{Prasad:1975kr}
  M.~K.~Prasad and C.~M.~Sommerfield,
  Phys.\ Rev.\ Lett.\  {\bf 35}, 760 (1975).




\bibitem{Kihara:2004yz}
  H.~Kihara, Y.~Hosotani and M.~Nitta,
  Phys.\ Rev.\ D {\bf 71}, 041701 (2005)
  [arXiv:hep-th/0408068].

\bibitem{abel:1829}
	N.~H.~Abel, 
  J.\ Reine Angew.\ {\bf 4}, (1829).


\bibitem{Murphy:1960}
  G.~M.~Murphy, 
  ``Ordinary Differential Equations and Their Solutions,''
  D. Van Nostrand Company, INC. USA





\end{thebibliography}
\end{document}